\documentclass[%
 reprint,
 amsmath,amssymb,
 aps,
]{revtex4-2}

\usepackage{graphicx}%
\usepackage{dcolumn}%
\usepackage{bm}%
 \usepackage{url}%
\usepackage{xcolor}

\begin{document}

\title{High-optical-depth, sub-Doppler-width absorption lines at telecom wavelengths\\ in hot, optically driven rubidium vapor }

\author{Inna Kviatkovsky$^1$, Lucas Pache$^1$, Viola-Antonella Zeilberger$^1$, Philipp~Schneeweiss$^1$, Jürgen Volz$^1$, Arno Rauschenbeutel$^1$, Leonid Yatsenko$^2$ }

\affiliation{$^1$ Institut f\"{u}r Physik, Humboldt-Universit\"{a}t zu Berlin, Unter den Linden 6,  Berlin 10099, Germany}

\affiliation{$^2$ Institute of Physics, National Academy of Sciences of Ukraine,    Nauki Avenue 46, Kyiv 03028, Ukraine}

\date{\today}

\begin{abstract}
	Doppler broadening presents a major limitation for high-resolution spectroscopy and nonlinear optics in room-temperature atomic vapors. Here, we demonstrate   the suppression of Doppler broadening accompanied by pronounced absorption on the upper transition of a three-level ladder system, achieved by dressing the intermediate state with a strong control field. As a concrete realization, we study a hot vapor of \(^{87}\)Rb  where the lower transition is driven by a strong control field resonant with the D$_2$ line at a wavelength of 780~nm, while a weak counterpropagating probe field at the telecom C-band wavelength of 1529~nm (\(5P_{3/2}\!\leftrightarrow\!4D_{5/2}\)) interrogates the dressed states.  We observe absorption features with a resonant optical depth  of approximately 4 and a full width at half maximum of about 17 MHz. Remarkably, this corresponds to an order-of-magnitude reduction relative to the Doppler width, while the optical depth on the upper transition of the ladder scheme exceeds that of the Doppler-broadened lower transition. The measured spectra are in good agreement with theoretical modeling. Combining high optical density with sub-Doppler-width absorption lines typically requires laser-cooled atoms, while our approach profits from the experimental simplicity of a hot-vapor platform.  
\end{abstract}

\maketitle


\section{\label{sec:intro}Introduction}

Atomic vapors are extensively employed as   experimental platforms for quantum applications, due to their accessibility and the technical simplicity of room-temperature operation \cite{davidson2023single, urvoy2015strongly, craddock2024high, kong2020measurement}. A fundamental limitation of such systems, however, is Doppler broadening of optical transitions induced by   the thermal velocity distribution of the atoms \cite{Foot2005AtomicPhysics, glorieux2018quantum}. This inhomogeneous broadening reduces spectral resolution and coherence, thereby substantially limiting the performance of a wide range of quantum-optical protocols \cite{hammerer2010quantum, lee2016highly, kash1999ultraslow, willis2009four, glorieux2023hot, finkelstein2023practical}.

Earlier studies~\cite{Svanberg1986,Svanberg:87}  demonstrated Doppler-free transmission effects in optically thick atomic media using two-level configurations, providing an important precursor to later developments in coherent spectroscopy of strongly absorbing vapors. 
In three-level systems, Doppler broadening can be suppressed by minimizing the wave-vector mismatch between the control and probe fields. In a $\Lambda$ configuration, copropagating beams with nearly degenerate wavelengths, typically separated only by hyperfine or Zeeman splittings, enable nearly complete Doppler cancellation \cite{harris1997electromagnetically,fleischhauer2005electromagnetically, finkelstein2023practical}. In ladder schemes, by contrast, sub-Doppler features can be obtained with counterpropagating beams, but the wave-vector matching condition is substantially more challenging to satisfy because the two optical transitions generally occur at different wavelengths \cite{urvoy2013optical, grove1995two, mohapatra2007coherent,gj9l-m7hh}. Several approaches have therefore been proposed to compensate for the wave-vector mismatch and recover sub-Doppler resolution \cite{reynaud1979experimental, finkelstein2019power, chang2007doubly}.

In this work, we introduce a Doppler-cancellation scheme tailored to ladder systems with strongly nondegenerate transitions. The lower transition is driven by a strong control field, while optimal Doppler cancellation is achieved by employing a weak probe field with a frequency approximately half that   of the lower-transition frequency, arranged   in a counterpropagating configuration. Importantly, this mechanism does not rely on velocity selection, in contrast to saturation absorption spectroscopy (SAS) \cite{haroche1972theory, letokhov1975nonlinear, martinez2015absolute}. Instead, it involves a broad range of atomic velocity classes, resulting in enhanced  optical depth (OD)  rather than a reduction in the number of participating atoms.

We investigate a specific ladder configuration in $^{87}$Rb, where  the control field drives the D$_2$ transition at a wavelength of 780~nm ($5S_{1/2}\!\leftrightarrow\!5P_{3/2}$), while a weak probe field addresses the  $5P_{3/2}\!\leftrightarrow\!4D_{5/2}$ transition at the   telecom C-band wavelength of 1529~nm.  The observed absorption features exhibit a linewidth reduced by approximately one order of magnitude relative to the Doppler width, with a full width at half maximum (FWHM) of about 17~MHz and an OD of approximately~4. We find good agreement between a dedicated numerical model and the experimental results. The chosen ladder scheme is particularly attractive because its upper transition lies in the technologically important telecom C-band. More generally, the proposed approach can be extended to other ladder systems with counterpropagating beams, with  optimal performance achieved when the ratio of the upper to lower transition frequencies is close to one half.

The simultaneous realization of large optical depth and narrow sub-Doppler resonances on upper optical transitions at telecom wavelengths represents the principal result of this work. This combination provides simultaneously high optical depth and narrow spectral response in a simple room-temperature atomic vapor, a regime that has remained difficult to access with conventional approaches. This unique capability opens new opportunities for quantum-optical and nonlinear-optical schemes operating at telecom wavelengths, including quantum-light generation, slow-light protocols, narrowband optical filtering, and coherent frequency conversion.

\begin{figure*} 
	\includegraphics[width=0.5\textwidth,angle=270]{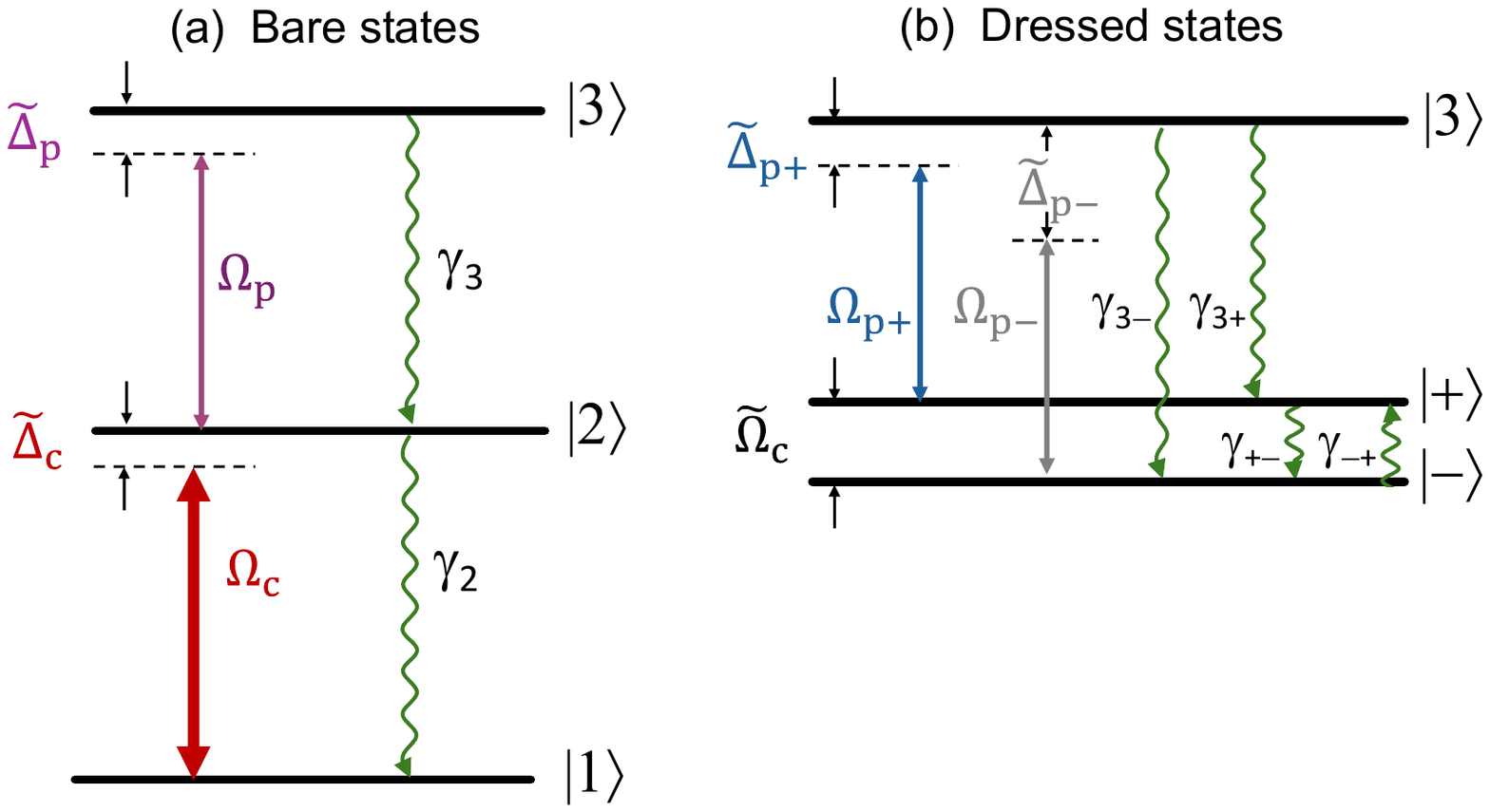}
	
	\caption{Excitation scheme of a ladder-type three-level atom in the bare-state (a) and dressed-state (b) representations.
		(a) A strong control field with frequency $\omega_c$, wave-vector magnitude $k_c$, and Rabi frequency $\Omega_c$ drives the transition between the initially populated ground state $|1\rangle$ and the intermediate state $|2\rangle$ with transition frequency $\omega_{12}$. In the atomic rest frame, the Doppler-shifted detuning is $\tilde{\Delta}_c=\Delta_c+k_c v_z$, where $\Delta_c=\omega_{12}-\omega_c$ is the detuning from the center of the Doppler-broadened transition and $v_z$ denotes the atomic velocity component along the  propagation direction of the control field.
		A weak probe field with frequency $\omega_p$, wave-vector magnitude $k_p$, and Rabi frequency $\Omega_p$ couples the transition $|2\rangle\!\leftrightarrow\!|3\rangle$  with Doppler-shifted detuning $\tilde{\Delta}_p=\Delta_p\pm k_p v_z$, where $\Delta_p=\omega_{23}-\omega_p$; the upper (lower) sign corresponds to copropagating (counterpropagating) probe and control fields.
		(b) In the dressed-state picture, the control field mixes states $|1\rangle$ and $|2\rangle$ into dressed states $|+\rangle$ and $|-\rangle$, separated in energy by $\hbar\tilde{\Omega}_c=\hbar\sqrt{\tilde{\Delta}_c^2+\Omega_c^2}$. The  dressed states are coupled to state $|3\rangle$ with effective Rabi frequencies $\Omega_{p,+}$ and $\Omega_{p,-}$ and detunings $\tilde{\Delta}_{p,+}$ and $\tilde{\Delta}_{p,-}$ [see Eqs.~(\ref{parameters1})--(\ref{parameters2})]. Wavy arrows denote incoherent spontaneous decay processes with rates $\gamma_2$ and $\gamma_3$.}
	
	\label{fig1:levels}
\end{figure*}

The paper is organized as follows. In Sec.~II, we analyze the emergence of strong sub-Doppler absorption within a simplified three-level model. Under the weak-probe approximation, we derive an analytical expression for the probe absorption coefficient and examine its behavior in the weak- and strong-saturation regimes of the control field. To clarify the physical origin of the Doppler-cancellation effect, we employ the dressed-state picture of atom-field interaction. Sections~III and~IV describe the experimental setup and the numerical model, respectively, including the Zeeman and hyperfine structure of $^{87}$Rb. In Sec.~V, we present the experimental observation of Doppler cancellation and compare the results with numerical simulations. Finally, Sec.~VI summarizes our findings and outlines future directions.

\section{\label{sec:simple}Simplified three-level model for strong sub-Doppler absorption}

We consider a gas of three-level atoms in a ladder configuration
[Fig.~\ref{fig1:levels}(a)] interacting collinearly with a strong control field with Rabi
frequency $\Omega_c$ and a weak probe field $\Omega_p$.
Collisions are neglected, so that the coherence decay rates are
$
\Gamma_{ij}=\frac{\gamma_i+\gamma_j}{2},
$
with $\gamma_1=0$, and $\gamma_{2,3}$ the spontaneous decay rates of the
excited states. The system is assumed to be nearly closed, with state
$|2\rangle$ decaying exclusively to $|1\rangle$, and the interaction
time exceeding all excited-state lifetimes.
We assume the Doppler-dominated regime
\(
k_c v_0,\, k_p v_0 \gg \gamma_2, \gamma_3,
\)
where $k_{c,p}=\omega_{c,p}/c$ and $v_0$ is the most probable atomic
velocity.

Unlike standard EIT schemes
\cite{urvoy2013optical, grove1995two, mohapatra2007coherent},
the control field with Rabi frequency $\Omega_c$ drives the transition
between the initially populated ground state $|1\rangle$ and the empty
excited state $|2\rangle$. As a result, the population in state $|2\rangle$ and coherence on the
$|1\rangle$–$|2\rangle$ transition are created only in the presence of
the control field; without it ($\Omega_c=0$), probe absorption on the
$|2\rangle$–$|3\rangle$ transition vanishes.

For our model, the expression for the absorption coefficient
$\alpha_p$
of a weak probe field
($\Omega_p\ll \gamma_i$) derived using the stationary solution of the density-matrix equations
reads:
\begin{widetext}
	\begin{align}
		\alpha_p = \alpha_{0p}
		\Re \!\int\! dv_z\, W(v_z)
		\frac{\Gamma_{23}\Omega_c^2/4}{
			\Bigl[
			\Gamma_{13}-i(\tilde{\Delta}_c+\tilde{\Delta}_p)
			+\dfrac{\Omega_c^2/4}{\Gamma_{23}-i\tilde{\Delta}_p}
			\Bigr]
			\Bigl[
			\tilde{\Delta}_c^2+\Gamma_{12}^2+\Omega_c^2/2
			\Bigr]},
		\label{absorption}
	\end{align}
\end{widetext}
where $\tilde{\Delta}_c=\Delta_c+k_c v_z$ and
$\tilde{\Delta}_p=\Delta_p\pm k_p v_z$ correspond to copropagating and
counterpropagating configurations of the probe and control lasers, respectively.
Here $\alpha_{0p}=\sigma_{23}N$,
$\sigma_{23}=\frac{3\lambda_{23}^2}{2\pi}\frac{\Gamma_{13}}{\Gamma_{23}}$,
and
$W(v_z)=(\sqrt{\pi}v_0)^{-1}e^{-v_z^2/v_0^2}$.

\subsection*{Weak-saturation regime}

\begin{figure} 
	\centerline{	\includegraphics[width=0.45\textwidth]{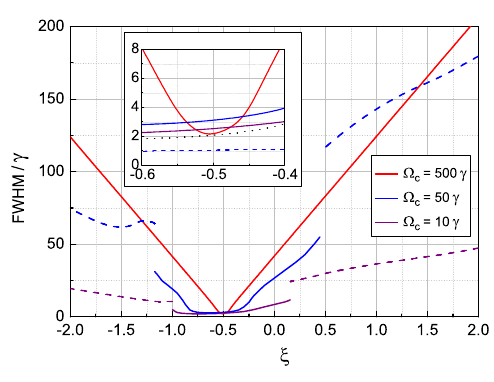}}

	\caption{Dependence of the full width at half maximum (FWHM) of a single Autler–Townes (AT) component on the ratio of the absolute values of the probe- and control-field wave vectors for three values of the control-field Rabi frequency $\Omega_c$, calculated for $\gamma_2=\gamma_3=\gamma$ and $k_c v_0 = 50\gamma$ (solid lines). The variable $\xi$ is defined as $\xi = k_p/k_c$ for copropagating probe and control fields and $\xi = -k_p/k_c$ for the counterpropagating configuration. For parameter regimes where the AT components are not spectrally resolved, the FWHM of the resulting single spectral feature is shown by the corresponding dashed lines. As a resolution criterion, we require that the signal at $\Delta_p = 0$ does not exceed one half of the maximum absorption signal, which occurs at $\Delta_p \simeq \Omega_c/2$. The inset shows the FWHM on an expanded scale; the dotted and dashed blue curves indicate, respectively, the inner and outer half-widths at half maximum (see definitions in Fig.~\ref{shape}) of an individual AT component for $\Omega_c = 50\gamma$.}

	\label{Fig:width}
\end{figure}

\begin{figure} 
	\includegraphics[width=\columnwidth]{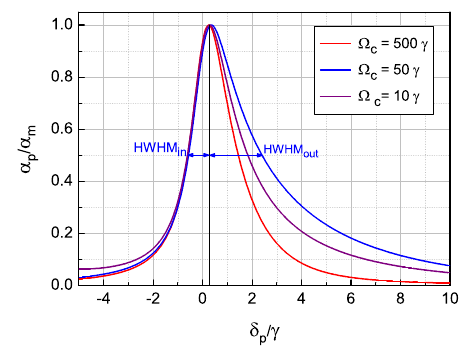}
	\caption{Shape of a single resolved  AT component: the probe absorption coefficient $\alpha_p$, normalized to its maximum value $\alpha_m$, as a function of the probe  detuning $\delta_p$ from the expected position of the AT component, $\delta_p=\Delta_p-\Omega_c/2$, for three values of the control-field Rabi frequency $\Omega_c$ ($\gamma_2=\gamma_3=\gamma$ and $k_c v_0=50\gamma$).}
	\label{shape}
\end{figure}
For $\Omega_c\ll\gamma_2$, the control field interacts only with atoms
within a narrow velocity group
$\Delta v_z\lesssim\gamma_2/k_c$ centered at
$v_{z,0}=-\Delta_c/k_c$.
The probe excites both one-photon $|2\rangle$–$|3\rangle$ transitions
and coherence-assisted two-photon $|1\rangle$–$|3\rangle$ transitions.
Their interference produces a Doppler-free Lorentzian feature in the probe absorption coefficient with
half-width
\begin{equation}
	\Gamma_\pm=\Gamma_{13}+\Bigl(1\pm\frac{k_p}{k_c}\Bigr)\Gamma_{12},
\end{equation}
centered at $\Delta_p=\pm(k_p/k_c)\Delta_c$ with peak absorption
\begin{equation}
	\alpha_p=\alpha_{0p}\frac{\sqrt{\pi}}{k_c v_0}
	\frac{\Gamma_{23}}{\Gamma_{12}\Gamma_\pm}\frac{\Omega_c^2}{4}.
\end{equation}
In this regime, the control field is unsaturated and decays along the propagation direction $z$ as
\begin{equation}
	\Omega_c^2(z)\propto e^{-\alpha_c z},
\end{equation}
where the linear absorption coefficient $\alpha_c$ is given by
\begin{equation}
	\alpha_c=\alpha_{0c}\,\frac{\Gamma_{12}\sqrt{\pi}}{k_c v_0}.
\end{equation}
with $\alpha_{0c}=\sigma_{12}N$  and
$
	\sigma_{12}=\frac{3\lambda_{12}^2}{2\pi}.
$
As a result, the characteristic length of the region in which atoms are efficiently driven by the control field  is of the order
$
	L_c\simeq \frac{1}{\alpha_c} 
$.
Consequently, the attainable OD on the excited-state transition $|2\rangle\text{--}|3\rangle$,
$
	OD(z)=\int_0^z \alpha_p(z')\,\mathrm{d}z',
$
cannot exceed
$
	OD_{\max}\simeq  {\alpha_p}/{\alpha_c}.
$
Within our model, this maximum OD reads
\begin{equation}
	OD_{\max}
	=\frac{k_c^2}{k_p^2}\,
	\frac{\Gamma_{13}}{\Gamma_{\pm}}\,
	I_{\mathrm{s }},
\end{equation}
where
$
	I_{\mathrm{s }}= {\Omega_c^2(z=0)}/{\gamma_2^2}
$
is the saturation parameter of the control field at the entrance of the absorbing medium.
Thus, in the weak-saturation regime $I_{\mathrm{s }}\ll 1$, the rapid attenuation of the control field  intrinsically restricts  the effective interaction length for probe absorption on the $|2\rangle\text{--}|3\rangle$ transition, and the resulting OD remains small, $OD\ll 1$, irrespective of further increases in atomic density or  length of the vapor cell.

\subsection*{Strong-saturation regime}

In the strong-saturation regime, $I_{\mathrm{s}} \gg 1$, the control-field intensity no longer decays exponentially with propagation distance but instead decreases approximately linearly. As a result, a significant population of the intermediate state is sustained over a much longer distance $L_c$, thereby enabling probe absorption over an extended interaction region.  
However, when $\Omega_c \gtrsim \Gamma_{12}$, Autler--Townes (AT) splitting emerges in the probe spectrum \cite{autler1955stark}, leading to the formation of a doublet structure. For Rabi frequencies exceeding a certain critical value $\Omega_{\mathrm{cr}}$, the single absorption peak observed in the weak-saturation regime splits into two components symmetrically positioned around $\Delta_p = 0$. The splitting, defined as the separation between the probe detunings $\Delta_p$ corresponding to the maxima of the two components, is approximately equal to the control-field Rabi frequency.       
 
\begin{figure}[t]
	\includegraphics[width= 1.2\columnwidth]{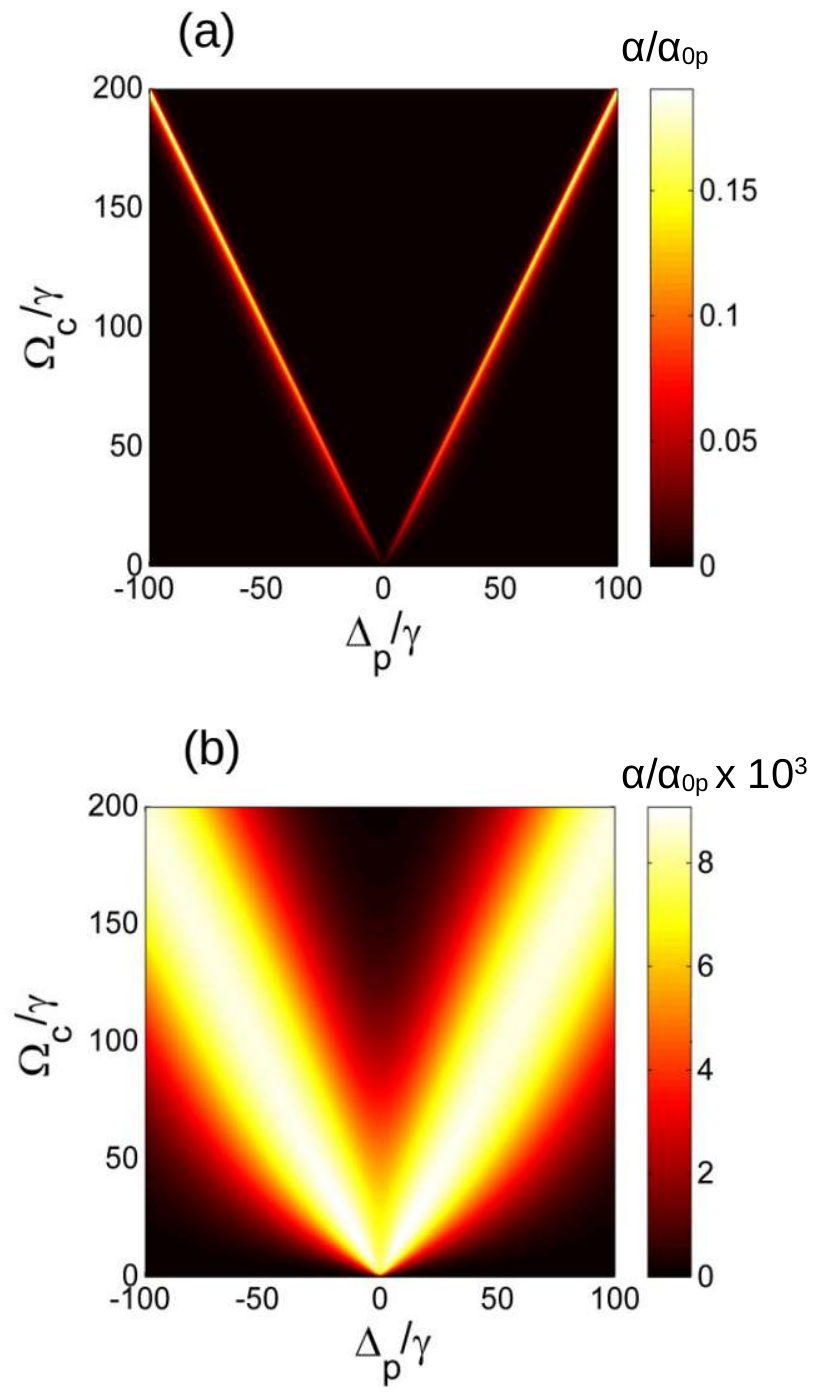}
	\vspace{-2.5cm}
\caption{Normalized probe absorption coefficient $\alpha/\alpha_{0p}$ as a function of the probe detuning $\Delta_p$ and the control-field Rabi frequency $\Omega_c$ for (a) counterpropagating and (b) copropagating configurations, calculated for $k_p=0.5k_c$, $\gamma_2=\gamma_3=\gamma$, and $k_c v_0=50\gamma$.}
	
	\label{2Dplots}
\end{figure}
The value of $\Omega_{\mathrm{cr}}$ and the spectral properties of the AT  doublet depend  strongly  on the ratio $k_p/k_c$ of the magnitudes of the collinear probe and control wave vectors.  Note that only collinear propagation is considered here. In a noncollinear configuration, unavoidable additional  Doppler broadening is present.  Figure~\ref{Fig:width} shows the  FWHM  of the AT doublet as a function of this ratio. Well-resolved sub-Doppler components occur predominantly for counterpropagating probe and control fields, with an optimal ratio \begin{equation}
 k_p/k_c =   0.5 .
 \label{condition}
\end{equation}
In the  vicinity  of this point (see inset of Fig.~\ref{Fig:width}), the FWHM remains below a few $\gamma$ for all considered values of the control-field Rabi frequency $\Omega_c$, indicating that Doppler broadening contributes only weakly to the linewidth.   
For intermediate control-field strengths, $\Omega_c \sim k_c v_0$,  as depicted by the blue line in Fig.~\ref{Fig:width},   Doppler  broadening  suppression persists over a fairly broad  range of wave-vector ratios, although the line shape of individual components   deviates from a Lorentzian. The doublet components exhibit pronounced asymmetry (see Fig.~\ref{shape}). In particular, markedly different half-widths at half maximum are observed on the outer side of the doublet and on the side adjacent to the neighboring AT component (see also the inset of Fig.~\ref{Fig:width}).
In the limit $\Omega_c \gg k_c v_0$,  as shown by the red line in Fig.~\ref{Fig:width},  the AT components become more symmetric and approach a Lorentzian line shape without Doppler broadening provided that condition~(\ref{condition})
is satisfied with higher accuracy than in the regime of the intermediate control-field strengths.

Figure~\ref{2Dplots} illustrates the Doppler-cancellation effect under the optimal condition~(\ref{condition}) over a broad range of control-field Rabi frequencies by direct comparison with the copropagating case where no such effect occurs.

\begin{figure}[t]
	\includegraphics[width=  \columnwidth]{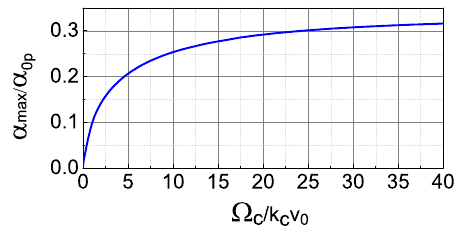}
	 
	\caption{Normalized peak absorption $\alpha_{\max}/\alpha_{0p}$ of a single AT component as a function of the control-field Rabi frequency $\Omega_c$, expressed in units of the Doppler width $k_c v_0$, for the counterpropagating configuration. The curves are calculated for $k_p=0.5k_c$, $\gamma_2=\gamma_3=\gamma$, and $k_c v_0=50\gamma$.}
	
	\label{abs_vs_omega_c}
\end{figure}
The normalized peak absorption $\alpha_{\max}/\alpha_{0p}$ of a single AT component as a function of the control-field Rabi frequency $\Omega_c$ is shown in Fig.~\ref{abs_vs_omega_c} over a wide parameter range. One observes a linear increase with $\Omega_c$ in the intermediate regime
\(
\gamma \lesssim \Omega_c \lesssim k_c v_0,
\)
which, in the strong-field limit $\Omega_c \gg k_c v_0$, is replaced by saturation of $\alpha_{\max}$ toward
\begin{equation}
	\alpha_{\max} = \alpha_{0p} \frac{\Gamma_{13}}{\Gamma_{13} + \Gamma_{23}}~.
	\label{max_abs}
\end{equation}
For $\Gamma_{13} \sim \Gamma_{23}$, this limiting value approaches the absorption expected for a Doppler-free medium with all atoms prepared in state $|2\rangle$.

\subsection*{Dressed states picture of Doppler cancellation  effect}

 The suppression of Doppler broadening observed in numerical simulations based on Eq.~(\ref{absorption}) can be readily understood within the dressed-state formalism.  Rather than working with the two ``bare'' states $|1\rangle$ and $|2\rangle$ coupled by the strong field $\Omega_c$, it is convenient to introduce the dressed-state basis
 \begin{align}
 	|\psi_- \rangle &= \cos\theta\,|1\rangle - \sin\theta\,|2\rangle, \\
 	|\psi_+ \rangle &= \sin\theta\,|1\rangle + \cos\theta\,|2\rangle,
 \end{align}
 where the mixing angle $\theta$ is defined by
 \begin{equation}
 	\tan 2\theta = \frac{\Omega_c}{\tilde{\Delta}_c}.
 \end{equation}
 The angle $\theta$ is chosen in the interval $0<\theta<\pi/2$, such that both $\sin\theta$ and $\cos\theta$ are positive. With this convention, the dressed state $|\psi_+\rangle$ always has a higher energy than $|\psi_-\rangle$. In the limit $\Omega_c\to 0$, the dressed states reduce to the bare states: $|\psi_+\rangle,|\psi_-\rangle \to |1\rangle,|2\rangle$ for $\Delta_c>0$, and $|\psi_+\rangle,|\psi_-\rangle \to -|2\rangle,|1\rangle$ for $\Delta_c<0$.
The energies of the dressed states $|\pm\rangle$ are given by
 \begin{equation}
 	\Delta E_\pm = \frac{\hbar}{2}
 	\left[
 	\tilde{\Delta}_c \pm \sqrt{\tilde{\Delta}_c^{\,2} + \Omega_c^{2}}
 	\right].
 \end{equation}
 In general, the two dressed states are   coupled due to the finite lifetime of the bare state $|2\rangle$, which induces a residual coherence between them. 
 However, in the regime of a large Rabi frequency $\Omega_c$, where the energy separation $\sqrt{\tilde{\Delta}_c^{\,2}+\Omega_c^{2}}$ greatly exceeds the decay rate $\gamma_2$, this coherence can be neglected. In this case, the interaction of the probe field with the atomic system on the transitions $|\pm\rangle \leftrightarrow |3\rangle$ can be treated as an interaction  with two independent effective two-level systems, giving rise to the two components of the AT doublet (see the excitation scheme for this case in Fig.~\ref{fig1:levels}(b)).
 
 To characterize the lineshape of each component, one needs to know the effective probe-field coupling $\Omega_p^{\pm}(v_z)$, the effective detuning $\tilde{\Delta}_p^{\pm}(v_z)$, the dressed-state populations $\rho^{\pm}(v_z)$, and the relaxation rate $\Gamma_{\pm,3}(v_z)$ of the coherence between states $|\pm\rangle$ and $|3\rangle$. All of these quantities depend explicitly on the atomic velocity $v$.
 For definiteness, we consider the effective two-level transition $|+\rangle\leftrightarrow|3\rangle$ and assume a resonant control field, $\Delta_c=0$. The relevant parameters then take the form
 \begin{align}
 	\Omega_{p,+}(v_z) &= \Omega_p \cos\theta, \label{parameters1}\\
 	\tilde{\Delta}_{p,+}(v_z) &= \tilde{\Delta}_p + \frac{1}{2}\tilde{\Delta}_c
 	- \frac{1}{2}\sqrt{\tilde{\Delta}_c^{\,2}+\Omega_c^{2}},
 	\label{detuning} \\
 	\Gamma_{+,3}(v_z) &= \gamma_3/2 + (\gamma_2/2)\cos^2\theta, \\
 	\rho_{+}(v_z) &= \frac{\sin^4\theta}{\sin^4\theta+\cos^4\theta}, \\
 	\tan 2\theta &= \frac{\Omega_c}{k_c v_z}.
 	\label{parameters2}
 \end{align}
 The linear absorption coefficient for this effective two-level system, with interaction parameters given by Eqs.~(\ref{parameters1})--(\ref{parameters2}), is then expressed as
 \begin{equation}
 	\alpha_p = \alpha_{0p}
 	\Re\int_{-\infty}^{\infty} d v_z\, W( v_z)\,
 	\cos^2\theta\,
 	\frac{ \gamma_3\,\rho^{+}( v_z)}
 	{\Gamma_{+,3}( v_z)-i\tilde{\Delta}_{p,+}( v_z)} .
 	\label{abs_dress}
 \end{equation}
This expression differs from the standard Voigt profile only through the unusual velocity dependence of the effective detuning. As follows from the definition~(\ref{detuning}) of the effective detuning $\tilde{\Delta}_{p,+}$, for counterpropagating probe and control fields, a resonant control field $\Delta_c=0$, and the specific wave-vector ratio $k_p=k_c/2$, the probe-field detuning takes the form
\begin{equation}
\tilde{\Delta}_{p,+}
=
\Delta_p-\frac{1}{2}\sqrt{k_c^2  v_z^2+\Omega_c^2}.
 \end{equation}
Thus, the linear dependence of the probe-field detuning on the atomic velocity cancels, while the remaining velocity dependence becomes quadratic in $v_z$. As a consequence, in the limit of a large control-field Rabi frequency, $\Omega_c\gg k_c v_0$, the Doppler broadening is completely negligible. Taking into account that for large $\Omega_c$ the mixing angle approaches $\theta=\pi/4$, one finds that the lineshape of each doublet component is purely Lorentzian, with a half width at half maximum (HWHM) equal to $\gamma_3/2+\gamma_2/4$, and with the peak absorption coefficient given by Eq.~(\ref{max_abs}).

While the suppression of Doppler broadening in the asymptotic regime ($\Omega_c \gg k_c v_0$) follows rather directly from the dressed-state resonance condition, the persistence of narrow resonances in the experimentally relevant intermediate regime ($\gamma_2,\gamma_3\ll\Omega_c\lesssim k_c v_0$) is considerably less obvious. In this regime, the range of atomic velocities contributing to the resonance is determined not   by    the one-dimensional Maxwell–Boltzmann distribution   $W( v_z)$, but predominantly by the velocity dependence of the coupling of the dressed state to the excited state, described by the factor $\cos^2\theta$ in Eq.~(\ref{abs_dress}), as well as by the velocity dependence of the population $\rho_{+}( v_z)$. As a result, the dominant contribution to the observed line originates from atoms with velocities
\[
| v_z|\lesssim \Omega_c/k_c < v_0 .
\]
Furthermore, for $k_p=k_c/2$, an atom with velocity $ v_z$ is detuned from the resonance   by
\begin{equation}
	\delta=\frac{k_c^2  v_z^2}{\Omega_c+\sqrt{\Omega_c^2+k_c^2  v_z^2}} .
\end{equation}
The corresponding  distribution $G(\delta)$      of effective detunings $\delta$ induced by the   Maxwell–Boltzmann velocity distribution $W(v_z)$ 
  is therefore strongly asymmetric: the probability density vanishes identically for negative detunings, $\delta<0$, while for $\delta>0$ it has a characteristic width of order $\Omega_c$. The resulting absorption profile, given by a convolution of a narrow Lorentzian of width $\Gamma_{+,3}\ll\Omega_c$ with the broad and asymmetric distribution $G(\delta)$, is itself asymmetric with respect to the nominal line center at $\Delta_p=\Omega_c/2$.  
Despite the fact that the distribution $G(\delta)$ associated with a single AT component has a width of order $\Omega_c$, the  HWHM  of this component is much smaller than $\Omega_c$. This reduction originates from the strong asymmetry of $G(\delta)$, for which $G(\delta<0)=0$. As a result, if we consider,   for definiteness, the component centered at  $\Delta_p=\Omega_c/2$, the wing of this component for $\Delta_p<\Omega_c/2$ is insensitive to Doppler broadening: the absorption profile remains Lorentzian, with the HWHM determined solely by the relaxation rate $\gamma$.
The   wing for $\Delta_p>\Omega_c/2$ of the AT component exhibits a larger HWHM; however, it is still governed by $\gamma$   owing to the dominant contribution of near-zero-velocity atoms.
 In the AT picture, atoms with small longitudinal velocities map onto effective detunings $\delta$ close to the AT resonance, producing a narrow contribution with a width of order $\gamma$. The number of such atoms scales as
\begin{equation}
	N_{\rm slow} \sim N\,\frac{\sqrt{\gamma\Omega_c}}{k_c v_0},
\end{equation}
which yields the contribution to the absorption coefficient at the AT-component center
\begin{equation}
	\alpha_{\rm slow} \sim \alpha_{0p}\,\frac{\sqrt{\gamma\Omega_c}}{k_c v_0}.
\end{equation}
In contrast, fast atoms with $k_c v_z \sim \Omega_c$     contribute near the AT-component center 
\begin{equation}
	\alpha_{\rm fast} \sim \alpha_{0p}\,\frac{\gamma}{k_c v_0},
\end{equation}
which is  much  smaller than $\alpha_{\rm slow}$ since $\gamma\ll\Omega_c$.
Consequently, the decrease of absorption with increasing detuning from the AT-component center is primarily due to the diminishing contribution of slow atoms. This explains why  neither Doppler nor power broadening determines the FWHM of an individual AT component. Instead, the width remains proportional to $\gamma$, differing only by a numerical prefactor.
Thus, in the intermediate regime $\gamma_2,\gamma_3 \ll \Omega_c \lesssim k_c v_0$, the asymmetric mapping between atomic velocities and effective detuning within a single AT component leads to an asymmetric resonance profile with respect to the maximum at $\Delta_p=\Omega_c/2$. Nevertheless, the HWHMs on both sides of the AT component are controlled by the relaxation rates $\gamma_i$, rather than by $\Omega_c$ or the Doppler width $k_c v_0$.

We emphasize that, in the experimentally relevant intermediate regime, where the narrowing is governed predominantly by the contribution of slow atoms, substantial sub-Doppler narrowing persists over a much broader range of wave-vector ratios than in the asymptotic regime of very large control-field Rabi frequencies. Consequently, under realistic operating conditions the proposed scheme does not require precise wavelength matching.

\begin{figure*}[t]
	\includegraphics[width= \textwidth]{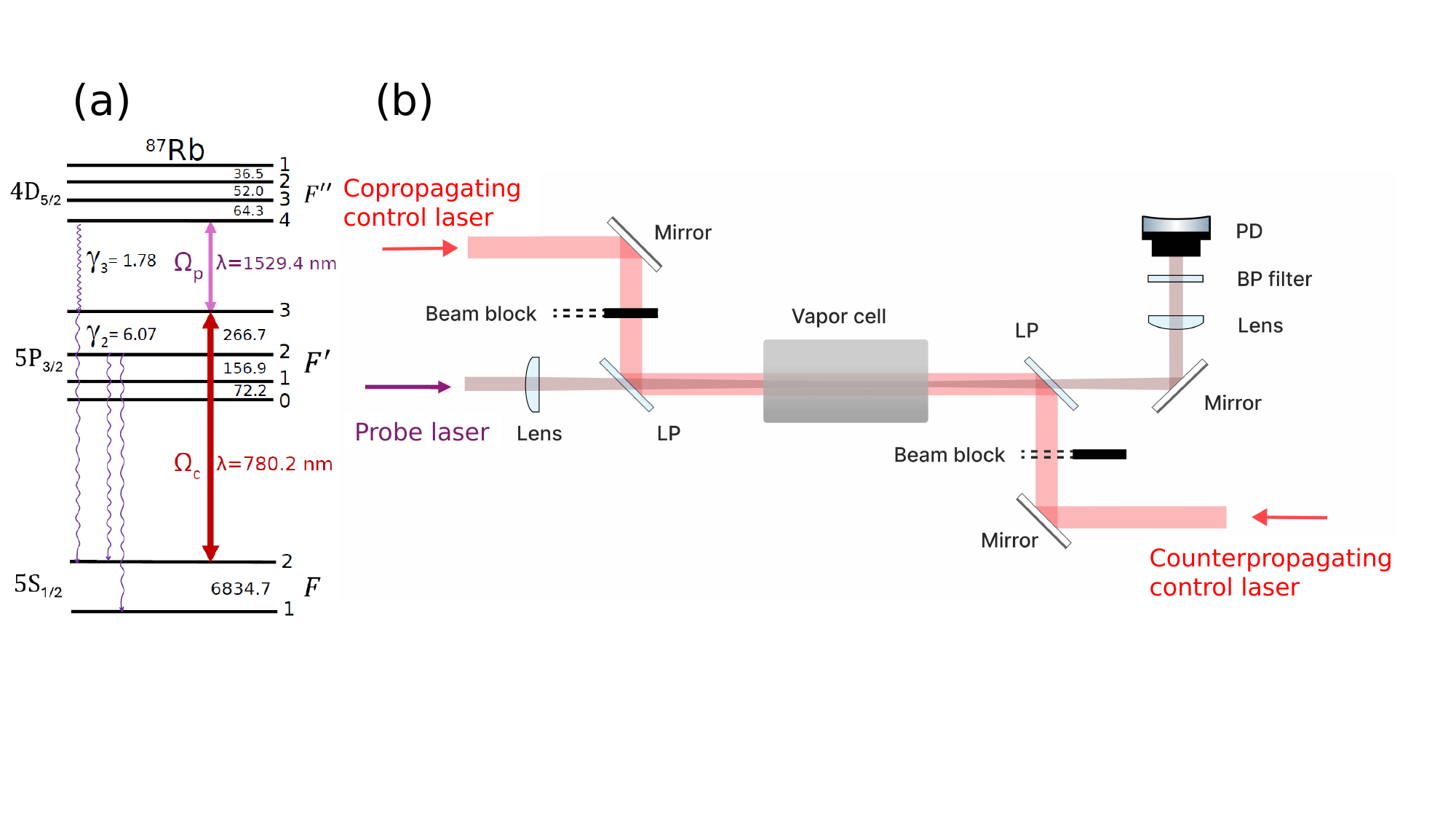}\vspace{-2.5cm}
\caption{
	(a) Hyperfine level scheme of $^{87}$Rb. Hyperfine intervals (in MHz) for the $5S_{1/2}$ and $5P_{3/2}$ levels are taken from Ref.~\cite{SteckRb87}, while those for the $4D_{5/2}$ level are taken from Ref.~\cite{PhysRevA.92.012501}. Spontaneous emission rates (in MHz) are taken from Ref.~\cite{PhysRevA.83.052508}. The control field that drives the lower transition is locked to the $F=2 \leftrightarrow F'=3$ transition (Doppler width at room temperature $k_c v_0 \approx 340$~MHz), while the probe field scans across the $F'=3 \leftrightarrow F''=4$ resonance ($k_p v_0 \approx 170$~MHz).
	(b) Schematic of the experimental setup. Probe light at a wavelength of 1529.4~nm is focused into the vapor cell, and the transmitted signal is detected by a photodiode (PD). Control light at a wavelength of 780.2~nm can be arranged to be either copropagating or counterpropagating with respect to the probe beam. Beam blocks are used to switch between copropagating and counterpropagating control--probe configurations. LP: long-pass filter; BP: band-pass filter.
}
	\label{fig:levels}
\end{figure*}

\section{\label{sec:setup}Experimental setup}

We investigate the interaction of two laser fields (control and probe) with \(^{87}\)Rb atoms in a ladder configuration involving the \(5S_{1/2}\)--\(5P_{3/2}\)--\(4D_{5/2}\) transitions. The corresponding hyperfine-level scheme is shown in Fig.~\ref{fig:levels}(a). The control laser, operating at a wavelength of 780.2~nm, drives the D$_2$ transition and is frequency-locked using a separate vapor cell to the cycling transition \(F=2 \leftrightarrow F'=3\). The probe laser, at a wavelength of 1529.4~nm, is scanned across the \(F'=3 \leftrightarrow F''=4\) transition.
For this configuration, the ratio of the wave-vector magnitudes is close to the optimal value for Doppler  cancellation,
\[
k_p/k_c = \lambda_c/\lambda_p \simeq 0.51.
\]

The control and probe beams are spatially overlapped in an uncoated rubidium vapor cell without buffer gas. An isotopically enriched $^{87}\mathrm{Rb}$ was used in the experiment. Besides reducing unwanted spectral overlap from $^{85}\mathrm{Rb}$, the use of an isotopically enriched $^{87}\mathrm{Rb}$ cell maximizes the fraction of atoms participating in the desired transition. This increases the attainable optical depth and correspondingly enhances the observed signal at a given vapor temperature. The cell has a length of 75~mm and a diameter of 19~mm and can be temperature controlled. Measurements were performed at room temperature ($t_C = 23.5\,^{\circ}\mathrm{C}$) and at an elevated temperature of $t_C = 38.5\,^{\circ}\mathrm{C}$. The control field can be applied from either direction, enabling copropagating or counterpropagating geometries relative to the probe beam [Fig.~\ref{fig:levels}(b)]. Beam blocks are used to select the propagation direction of the control field, enabling direct comparison between the two configurations. No magnetic shielding or active compensation of residual magnetic fields was required under our experimental conditions, since the MHz-scale absorption features investigated here are only weakly affected by typical ambient magnetic fields.

During the probe-frequency scan, the transmitted probe power is recorded with a variable-gain amplified photodiode. The experimental OD is obtained by normalizing the probe transmission measured in the presence of the control field ($\mathrm{P}_{\mathrm{sig}}$) to that recorded in its absence ($\mathrm{P}_{\mathrm{ref}}$), which serves as a reference. The dark current of the photodiode is subtracted from both signals. Because both signals experience identical optical losses along the detection path, these losses cancel upon normalization, ensuring that the extracted absorption reflects only the contribution of the atomic ensemble. The transmission is then converted to OD according to
\begin{equation}
	 {OD} = -\ln\!\left(\frac{ {P}_{\mathrm{sig}}}{ {P}_{\mathrm{ref}}}\right).
\end{equation}

The waist radii of the counterpropagating and copropagating control beams at the center of the cell are $905(3)\,\mu\mathrm{m}$ and $888(6)\,\mu\mathrm{m}$, respectively, while the probe-beam waist radius is $128(4)\,\mu\mathrm{m}$. Both the probe and control beams were spatially filtered through single-mode fibers before entering the vapor cell, resulting in well-defined Gaussian input profiles. The control beam is deliberately chosen to be substantially wider than the probe beam to ensure that atoms interacting with the probe experience an approximately uniform control-field Rabi frequency across the probed region. All optical fields are linearly polarized with mutually parallel polarization axes.

 \section{\label{sec:numerics}Numerical model}

  The actual ladder system in $^{87}$Rb studied in this work is significantly more complex than the idealized three-level model analyzed in Sec.~\ref{sec:simple}, which we introduced solely to elucidate the basic mechanism of Doppler cancellation in a ladder configuration. To accurately predict the properties of the narrow spectral features generated at telecom wavelengths and to enable a quantitative comparison with the experimental data, we developed a comprehensive numerical model for calculating the OD of an $^{87}$Rb vapor driven by a strong control field and interrogated by a weak probe field.

 The model explicitly accounts for Doppler broadening, the multilevel hyperfine structure of the atom, Zeeman sublevels, optical pumping effects, and the spatial evolution of the optical fields during propagation. This approach provides a direct link between the microscopic atomic response and the macroscopic absorption spectra measured in the experiment.

\subsection*{Propagation of optical fields}

 The probe field at the entrance of the vapor cell is modeled as a Gaussian beam whose wavefront curvature is chosen so that the beam waist lies at the center of the cell in the absence of the control field.   Its propagation through the atomic medium is described by numerically solving the scalar wave equation in the paraxial approximation for a medium characterized by a spatially inhomogeneous complex absorption coefficient. The probe field is assumed to be sufficiently weak so that saturation effects can be neglected.

The control field is treated self-consistently within the same theoretical framework. At the entrance of the cell, it is modeled as a Gaussian beam focused at the cell center, and its propagation is calculated within the paraxial approximation, including the effects of absorption and dispersion. Due to the large OD of the atomic medium at the control-field wavelength, the beam undergoes significant spatial reshaping as it propagates through the vapor.

In particular, the model naturally incorporates nonlinear effects such as nonlinear diaphragming arising from radially inhomogeneous saturation of the absorption, as well as nonlinear focusing and defocusing due to the radial variation of the refractive index induced by the spatially varying control-field intensity. These effects play a crucial role in determining the spatial structure of the control field inside the vapor cell and, consequently, the local Rabi frequency 
$\Omega_c(r,z)$ and the probe field experienced by the atoms in the interaction region.
   
Thus, the theoretical model does not assume fixed Gaussian beam profiles inside the cell: only the input fields are taken to be Gaussian, whereas their diffraction and transverse reshaping during propagation through the spatially inhomogeneous atomic medium are calculated self-consistently.
\subsection*{Atomic response and absorption coefficients}

The local absorption coefficients for both the probe and control fields are calculated from the stationary solution of the density-matrix equations for a $^{87}$Rb atom interacting with a strong control field and a weak probe field. The calculation fully accounts for the hyperfine structure of the $5S_{1/2}$ and $5P_{3/2}$ levels, including all magnetic sublevels. The optical fields are assumed to be linearly polarized.

A key simplifying assumption of the model is the absence of an external magnetic field. As a result, coherences between states with different magnetic quantum numbers $M$ are neglected. Nevertheless, the population distribution among the magnetic sublevels becomes strongly nonuniform due to optical pumping. This redistribution is explicitly included by incorporating spontaneous decay rates between magnetic sublevels, determined by the appropriate Clebsch--Gordan coefficients.

The relaxation of the populations of the $2F+1$ magnetic sublevels of the ground states $F=1$ and $F=2$ is described phenomenologically by a single rate $\gamma_{\rm tr}$ associated with atomic transit through the laser beam. The relaxation term for the density-matrix elements $\rho_{F,M;F,M}$ is written as
\begin{equation}
	\left.\frac{d\rho_{F,M;F,M}}{dt}\right|_{\rm rel}
	=
	-\gamma_{\rm tr}\left(\rho_{F,M;F,M}-\rho_{F,M;F,M}^{(0)}\right),
\end{equation}
where $\gamma_{\rm tr}=1/\tau_0$, with $\tau_0$ the mean transit time across the Gaussian beam. Since the ground-state hyperfine splitting is much smaller than the thermal energy $kT$, the equilibrium populations are independent of $F$ and $M$, giving $\rho_{F,M;F,M}^{(0)}=1/8$.
However, in the presence of a strong control field, the assumption of equal populations of the $F=1$ and $F=2$ states must be revised. The control field is resonant with the $F=2\leftrightarrow F'=3$ transition, from which spontaneous decay returns exclusively to $F=2$. At sufficiently high control-field intensities, however, off-resonant excitation of the $F'=2$ level becomes significant, enabling decay into the $F=1$ ground-state manifold and resulting in efficient optical pumping of the medium.

In low-density vapor cells with small OD, atoms optically pumped during a single transit typically relax back to thermal equilibrium after wall collisions before re-entering the interaction region. In contrast, at large OD and high control-field powers, atoms may not fully relax between successive transits. In addition, photons scattered by atoms in the interaction region can be reabsorbed by atoms approaching the beam, leading to a global optical orientation of the atomic ensemble even though the direct light--atom interaction occurs only in a spatially limited region.
To account for this effect, we introduce a phenomenological description of optical pumping by allowing the equilibrium populations $\rho_{2,M;2,M}^{(0)}$ of the magnetic sublevels of the $F=2$ state to depend on the control-field power,
\begin{equation}
	\rho_{2,M;2,M}^{(0)}
	=
	\frac{1}{8}\,
	\frac{1}{\sqrt{1+P_c/P_{\rm op}}},
	\label{opt_pump}
\end{equation}
where $P_c$ is the control-field power and $P_{\rm op}$ is a phenomenological parameter. Its value is determined by matching the numerical results to the experimental data at high control powers, yielding $P_{\rm op}\approx 20~\mathrm{mW}$. This phenomenological description captures the essential redistribution of populations among the ground-state hyperfine levels observed experimentally at large optical depth and strong control fields.

\subsection*{Doppler averaging and numerical implementation}

 Another important simplifying assumption is the neglect of modifications to the atomic velocity distribution caused by momentum transfer from the light fields.
  This is justified because the relevant processes occur on very different time scales. Optical pumping can create long-lived nonequilibrium spin populations, but any disturbances in the atomic velocities relax much more quickly, typically between successive passes of the atoms through the interaction region. As a result,   when atoms return to the interaction region, their velocities are again well described by a standard Maxwellian distribution. 
  Under these assumptions, we numerically solve the stationary algebraic system of density-matrix equations for an atom with a given longitudinal velocity $v_z$. Using the resulting optical coherences, we compute the absorption coefficient at the corresponding Doppler-shifted detuning. The total absorption coefficient is obtained by averaging over the thermal velocity distribution.
Since the optical fields evolve during propagation, the absorption coefficient must be recalculated at each propagation step. To significantly accelerate the calculations, we employ interpolation over a precomputed lookup table of absorption coefficients calculated for a discrete set of Rabi frequencies. The same procedure is applied to both the control and probe fields.
Within this framework, the suppression of Doppler broadening arises naturally from the velocity-dependent atomic response in the presence of the strong control field. The numerical model thus enables identification of the atomic velocity classes responsible for the observed sub-Doppler and Doppler-free features and provides a quantitative link between the microscopic density-matrix dynamics and the macroscopic absorption spectra. Additional details of the numerical model and the simulation codes used in this work are available in the publicly accessible repository \cite{Codes}.

 \section{\label{sec:exp}Experimental Results and Comparison with Theory}

 \begin{figure*}
 	\includegraphics[width=0.7\textwidth]{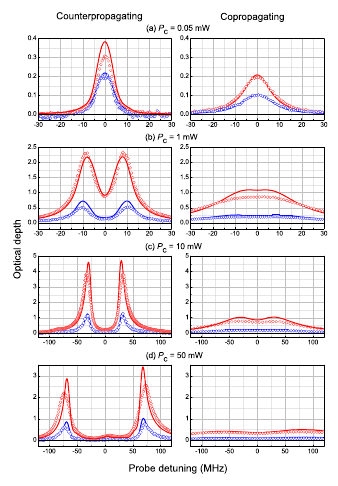}
 	\caption{Optical depth of a $^{87}$Rb vapor cell for the probe field at a wavelength of 1529.4~nm as a function of the probe detuning from the $F'=3  \leftrightarrow  F''=4$ transition. The left (right) panel corresponds to the counterpropagating (copropagating) configuration. Data are shown for four control-field powers $P_c = 0.05$, 1, 10, and 50~mW. Experimental data are shown as circles (blue: room temperature, $t_C = 23.5\,^{\circ}\mathrm{C}$; red: heated cell, $t_C = 38.5\,^{\circ}\mathrm{C}$), while solid lines of the corresponding colors represent numerical simulations.}
 	\label{fig:exp_res}
 \end{figure*}
 
 In this section we present a quantitative comparison between experimental transmission spectra and numerical simulations for copropagating and counterpropagating geometries.
 Figure~\ref{fig:exp_res} shows the OD for the probe field as a function of  its  detuning from the $F'=3 \leftrightarrow F''=4$ transition for copropagating and counterpropagating beam configurations, measured at four values of the control-field power.  
 
  At low control power ($P_c = 50~\mu$W), the absorption profiles for the two configurations are very similar. The main difference is that the resonance in the counterpropagating configuration is approximately 1.5 times narrower than that in the copropagating case. It is worth noting that, in the limit of very low control power, the simple three-level model predicts a linewidth ratio of $\Gamma_+/\Gamma_- \approx 2.3$ between the copropagating and counterpropagating geometries (see Eq. (2)). However, at a control-laser power of $50~\mu$W, the Autler–Townes splitting becomes comparable to the resonance linewidth and thus remains insufficient to fully resolve the two components. This, combined with transit-time broadening, reduces the experimentally observed linewidth ratio to approximately 1.5. 
    
With increasing control power, the counterpropagating configuration exhibits a  resolved AT doublet, starting from $P_c \approx 1$~mW. The well-resolved  individual components of the doublet (for $P_c > 5$~mW)  are   asymmetric, as predicted by the simplified three-level model. The maximum measured optical depth, $\mathrm{OD} \simeq 4$, is reached at $P_c \approx 10$~mW. In this regime, a pronounced enhancement of the OD with increasing cell temperature is observed. In contrast, for the copropagating configuration the absorption profiles broaden significantly as the control power increases, while the peak OD remains small, reaching only $\mathrm{OD} \simeq 1$ at $P_c = 10$~mW.
 
\begin{figure}
 	\includegraphics[width=\columnwidth]{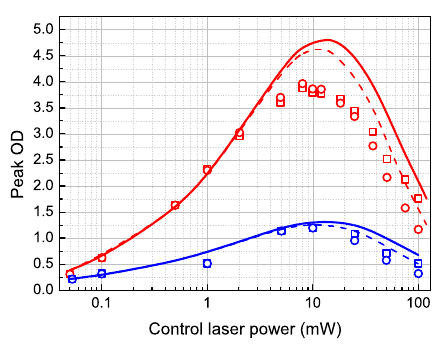}
 \caption{Peak OD for the counterpropagating control--probe configuration as a function of the control laser power, measured at two different cell temperatures (blue: room temperature, $t_C = 23.5\,^{\circ}\mathrm{C}$; red: heated cell, $t_C = 38.5\,^{\circ}\mathrm{C}$). Solid (dashed) lines show numerical simulations, while squares (circles) represent experimental data for the right (left) components of the AT doublet.}
 	\label{fig:OD_vs_Pc}
 \end{figure}

\begin{figure}[b]
 	\includegraphics[width=\columnwidth]{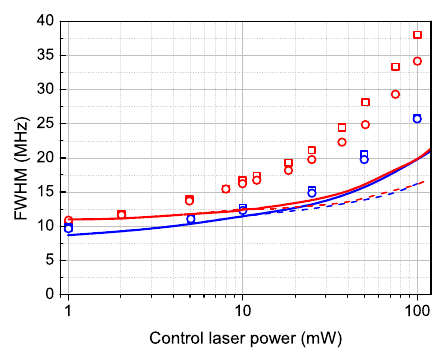}
 	\caption{ FWHM  of the AT doublet components for the counterpropagating control--probe configuration as a function of the control laser power. The notation is the same as in Fig.~\ref{fig:OD_vs_Pc}.}
 	\label{fig:hwhm}
 \end{figure}

 The theoretically predicted OD was calculated using the model described in the previous section with a phenomenological optical pumping parameter $P_{\mathrm{op}} = 20$~mW. With this choice, optical pumping significantly affects the calculated spectra only at the highest control power, $P_c = 50$~mW, shown in Fig.~\ref{fig:exp_res}. Overall, the numerical simulations reproduce the main experimental trends for all configurations considered.
  Figure~\ref{fig:OD_vs_Pc} shows the peak OD for the counterpropagating control--probe configuration as a function of the control-laser power, measured at room temperature and for a moderately heated cell. The experimental values of the peak OD, as well as the FWHM of the AT components, were obtained by fitting the measured OD spectra as a function of the probe detuning $\Delta_p$ with high-order polynomials. This procedure yields smooth dependences of OD on detuning, from which the frequency position of the maximum, the peak OD, and the FWHM were determined numerically. Although the individual AT components are generally asymmetric, we characterize their widths by the corresponding HWHM (or FWHM) values as a consistent measure of spectral extent; the conclusion that these resonances remain substantially narrower than the Doppler-broadened profile is independent of the particular linewidth convention employed. The statistical uncertainties of OD and FWHM are smaller than the size of the symbols used to represent the experimental data in Figs.~\ref{fig:OD_vs_Pc} and \ref{fig:hwhm}.  
This behavior can be attributed primarily to optical pumping  of atoms into the uncoupled ground state $F=1$ (see also Eq.~\ref{opt_pump}). 
  In contrast to the predictions of the simplified three-level model (see Fig.~\ref{abs_vs_omega_c}), the peak OD exhibits a pronounced maximum at a control power of approximately 10~mW. This power corresponds to an AT splitting of about 60~MHz, which is much smaller than the Doppler width, $k_c v_0 = 340$~MHz. Upon further increase of the control power, the peak OD decreases rather than saturating at a constant level, as predicted by the simplified model. This behavior can be attributed primarily to optical pumping  of atoms into the uncoupled ground state $F=1$ (see also Eq.~\ref{opt_pump}). 
A comparison between the experimental data and the numerical calculations presented in Figs.~\ref{fig:exp_res} and~\ref{fig:OD_vs_Pc} shows good overall agreement. The discrepancies observed predominantly at large optical depths can plausibly be attributed, at least in part, to the relatively strong probe field used in the experiment, whereas the model assumes a vanishing saturation parameter for the probe beam.
The room-temperature data were obtained with a probe power $P_p = 70.8(1.4)\,\mathrm{nW}$. In contrast, to maintain an acceptable signal-to-noise ratio at high optical depth ($\mathrm{OD} \sim 4$), the probe power in the experiment with the heated cell was chosen to be $P_p = 351(2)\,\mathrm{nW}$, which is comparable to the saturation power $P_{\mathrm{sat}} = 770(90)\,\mathrm{nW}$. The latter value was determined experimentally using the relation
\begin{equation}
	OD_{\max}(P_p) = OD_{\max}(0)\left(1 - \frac{P_p}{P_{\mathrm{sat}}}\right),
\end{equation}
where $OD_{\max}(P_p)$ denotes the peak optical depth of a single component of the AT doublet, valid for $P_p \ll P_{\mathrm{sat}}$. For realistic values $P_p / P_{\mathrm{sat}} \approx 0.2\text{--}0.5$, the measured peak OD is therefore expected to be reduced by several tens of percent compared with the value corresponding to an ideally weak, non-saturating probe.
The residual discrepancies   may arise also from a combination of factors, including uncompensated residual magnetic fields, the simplified treatment of ground-state population relaxation in the model, and uncontrolled reflections of the laser beams at the cell windows, as well as scattering of the probe beam from optical elements, which can lead to a small fraction of the probe light reaching the photodetector without interacting with the vapor.

 The  FWHM  of the absorption features is plotted in Fig.~\ref{fig:hwhm} as a function of the control power. Both calculated and experimental FWHM values increase with increasing control power; however, the observed broadening remains well below the Doppler width. The dominant contribution to this increase arises from the inhomogeneous axial intensity distribution of the control field inside the cell due to its absorption. This spatial variation leads to inhomogeneous broadening of the spectral components, since the frequency position the maximum of  each component   is determined by the local Rabi frequency of the control field.
 The remaining deviations between the experimental and calculated FWHM values are most likely associated with additional power broadening caused by the relatively high probe power $P_p$, an effect that is not included in the numerical model.

\section{\label{sec:discus}Discussion and Conclusions}

From the perspective of using rubidium vapor to generate narrow and strongly absorbing spectral features in the telecom C-band range, the present work demonstrates promising performance in the counterpropagating configuration at a control laser power of approximately $10$~mW. Under these conditions, at room temperature, the peak OD reaches values slightly above $1$, with a  FWHM  of about $11$~MHz. For a moderately heated cell at $38.5\,^{\circ}\mathrm{C}$, the peak OD increases to values around $4$, while the FWHM remains relatively small, approximately $17$~MHz. At the same control power, the copropagating configuration exhibits strong broadening of the absorption feature, approaching the Doppler width. In this case, the maximum OD remains low, reaching only about $0.3$ at room temperature and approximately $1$ for the heated cell. These observations clearly demonstrate the advantage of the counterpropagating geometry for achieving narrow and highly absorbing spectral features.

\begin{figure}
\includegraphics[width=\columnwidth]{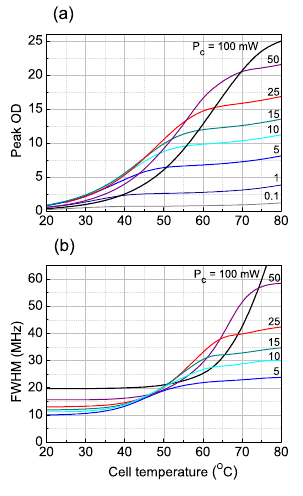}
	\caption{Calculated temperature dependence of the peak OD (panel (a)) and the FWHM (panel (b)) of the right component of the AT doublet for different values of the control-laser power. The FWHM is shown only for control powers for which the AT doublet is well resolved.}
	\label{fig:OD_vs_tc}
\end{figure}

For many prospective quantum applications based on hot atomic vapors, OD significantly larger than those demonstrated here are required. Since the developed numerical model provides a quantitative description of the experimental observations, it can be employed to predict the achievable OD in regimes of higher cell temperature, where the OD is expected to exceed the limits accessible with the current experimental setup.
Figure~\ref{fig:OD_vs_tc}(a) shows the calculated temperature dependence of the peak OD of the right component of the AT doublet for several values of the control-laser power. The FWHM as a function of temperature is shown in Fig.~\ref{fig:OD_vs_tc}(b) for the case of a well-resolved AT doublet. The simulations are performed using the geometrical parameters of the vapor cell and the laser beams employed in the experiment. An increase in optical depth could also, in principle, be achieved by increasing the interaction length, for example, by using a longer vapor cell or concatenated cells. However, in the presence of strong propagation effects and radially inhomogeneous saturation, changing the cell length is not generally equivalent to changing the atomic density at a constant interaction length. In the present experiment, varying the vapor temperature therefore provides a simpler and more practical control parameter.
For a given control-laser power, increasing the temperature initially leads to an increase in the OD, consistent with the growth of the saturated rubidium vapor pressure. However, beyond a certain temperature, stronger absorption of the control field reduces its intensity during propagation inside the cell to values close to the saturation power before reaching the end of the cell. After this point, exponential attenuation of the control field sets in. As a result, the control field becomes almost completely absorbed within the cell, and further heating no longer leads to an increase in the probe-field OD. As a consequence, the AT splitting becomes increasingly position dependent along the propagation direction. This spatial inhomogeneity leads to an effective broadening of the observed probe resonance, even while the optical depth continues to increase. The resulting intermediate propagation regime therefore represents a compromise between maximizing the optical depth and maintaining minimal linewidth.  
Importantly, as seen in Fig.~\ref{fig:OD_vs_tc}(a), this effect does not fundamentally limit the achievable OD when control-laser powers in the range of 10--25~mW are used. In this regime, the model predicts OD values exceeding 10 and approaching 15. Increasing the OD from 10 to 15 is accompanied by an increase of the FWHM from about 20~MHz to 30--35~MHz, depending on the control-laser power. These estimates indicate that the proposed method for generating highly absorbing and narrow spectral features can be extended well beyond the experimentally accessible OD range demonstrated here. Such narrow, high-OD absorption lines are promising for the realization of quantum light sources based on antibunching effects previously predicted and demonstrated in cold-atom systems \cite{PhysRevLett.127.123602,10.21468/SciPostPhysCore.6.2.041}.

In summary, we have demonstrated a Doppler-cancellation approach for strongly nondegenerate  wavelengths in a ladder-type excitation scheme. In this approach, a strong control field dresses the lower transition, resulting in an optimally narrow absorption feature when the wavelength ratio satisfies $k_c/k_p = 2$. Starting from an idealized three-level model, we identified three distinct operating regimes corresponding to strong ($\Omega_c \gg k_c v_0$), intermediate, and weak ($\Omega_c \ll \gamma_2$) control fields. We experimentally implemented this scheme in $^{87}$Rb, where the upper transition at 1529~nm lies in the telecom band and is therefore of direct relevance for communication applications.
Due to the hyperfine structure of the real atomic system, the best performance is achieved in the intermediate control-field regime, allowing operation with moderate laser powers. Under these conditions, we observe peak OD values exceeding 4 with linewidths only about five times the natural linewidth, corresponding to roughly a tenfold reduction in linewidth compared with the inhomogeneously Doppler-broadened profile. Crucially, such large OD values cannot be obtained by simply heating the vapor cell and instead constitute a direct manifestation of the Doppler-cancellation mechanism discussed here. In this regime, a broad range of atomic velocity classes contributes to absorption under nearly identical resonance conditions. As a result, the absorption on the upper transition of the ladder system becomes both significantly narrower than the Doppler width and stronger than that of the Doppler-broadened lower transition.

 Finally, we developed a dedicated numerical model that incorporates the main broadening mechanisms arising from deviations from an ideal three-level system. The experimental observations show good agreement with the model predictions. 
  Upon heating the cell, a further increase in OD is expected without significant additional broadening.

The presented approach enables the realization of narrow atomic absorption features while retaining the simplicity and practicality of room-temperature atomic vapors. We emphasize that the proposed scheme does not aim at realizing a conventional frequency reference. Rather, its key advantage is the simultaneous realization of narrow spectral features on upper transitions and large optical depth in an atomic ladder scheme. We have shown that this combination can be achieved at telecom wavelengths in hot atomic vapors, opening new opportunities for coherent and nonlinear light–matter interactions in a technologically important spectral region.

\acknowledgments
  We acknowledge funding by the Alexander von Humboldt Foundation in the framework of the Alexander von Humboldt Professorship endowed by the Federal Ministry of Education and Research, as well as direct funding by the Federal Ministry of Education and Research (Collaborative Research Project NetiQueT).

\bibliography{apssamp} 

\end{document}